\documentclass[letterpaper,twocolumn,aps,pra,floatfix,showpacs,tightenlines,superscriptaddress]{revtex4-1}
\usepackage[T1]{fontenc}
\usepackage[utf8]{inputenc}
\usepackage{color}
\usepackage{amsmath}
\usepackage{amssymb}
\usepackage{graphicx}
\usepackage{esint}
\usepackage[unicode=true,
 bookmarks=false,
 breaklinks=false,pdfborder={0 0 1},backref=section,colorlinks=false]
 {hyperref}
\usepackage{breakurl}

\makeatletter

\@ifundefined{textcolor}{}
{%
 \definecolor{BLACK}{gray}{0}
 \definecolor{WHITE}{gray}{1}
 \definecolor{RED}{rgb}{1,0,0}
 \definecolor{GREEN}{rgb}{0,1,0}
 \definecolor{BLUE}{rgb}{0,0,1}
 \definecolor{CYAN}{cmyk}{1,0,0,0}
 \definecolor{MAGENTA}{cmyk}{0,1,0,0}
 \definecolor{YELLOW}{cmyk}{0,0,1,0}
}

\usepackage{times}

\newcommand{\hc}{\mathrm{H.c.}}
\newcommand{\tr}{\mathrm{tr}}

\newcommand{\1}{\leavevmode{\rm 1\ifmmode\mkern  -4.8mu\else\kern -.3em\fi I}}

\makeatother

\begin{document}

\title{Small quench dynamics as a probe for trapped ultracold atoms}

\author{Sunil Yeshwanth}

\affiliation{Department of Physics \& Astronomy and Center for Quantum Information
Science \& Technology, University of Southern California, Los Angeles,
CA 90089-0484, USA}

\author{Marcos Rigol}

\affiliation{Department of Physics, The Pennsylvania State University, University
Park, PA 16802, USA}

\author{Stephan Haas}

\affiliation{Department of Physics \& Astronomy and Center for Quantum Information
Science \& Technology, University of Southern California, Los Angeles,
CA 90089-0484, USA}

\author{Lorenzo Campos Venuti}

\affiliation{Department of Physics \& Astronomy and Center for Quantum Information
Science \& Technology, University of Southern California, Los Angeles,
CA 90089-0484, USA}
\begin{abstract}
Finite systems of bosons and/or fermions described by the Hubbard
model can be realized using ultracold atoms confined in optical lattices.
The ground states of these systems often exhibit a coexistence of
compressible superfluid and incompressible Mott insulating regimes.
We analyze such systems by studying the out-of-equilibrium dynamics
following a weak sudden quench of the trapping potential. In particular,
we show how the temporal variance of the site occupations reveals
the location of spatial boundaries between compressible and incompressible
regions. The feasibility of this approach is demonstrated for several
models using numerical simulations. We first consider integrable systems,
hard-core bosons (spinless fermions) confined by a harmonic potential,
where space separated Mott and superfluid phases coexist. Then, we
analyze a nonintegrable system, a $J$-$V$-$V'$ model with coexisting
charge density wave and superfluid phases. We find that the temporal
variance of the site occupations is a more effective measure than
other standard indicators of phase boundaries such as a local compressibility.
Finally, in order to make contact with experiments, we propose a \emph{consistent
estimator} for such temporal variance. Our numerical experiments show
that the phase boundary is correctly spotted using as little as 30
measurements. Based on these results, we argue that analyzing temporal
fluctuations is a valuable experimental tool for exploring phase boundaries
in trapped atom systems. 
\end{abstract}


\maketitle

\section{Introduction}

The collective behavior of ultracold atoms in optical lattices can
be tuned by varying the depth of lattice potentials, thus adjusting
the ratio between the strength of the on-site interaction $U$ and
the hopping parameter $J$. In this manner, a quantum phase transition
between a superfluid (shallow lattice) and a Mott insulator (deep
lattice) can be induced \citep{greiner_quantum_2002,stoferle_moritz_04,spielman_phillips_07}.
An important feature in those experiments is the presence of a (to
a good approximation) harmonic trap, which results in the coexistence
of superfluid and Mott domains for a wide range of values of $U/J$
\citep{batrouni_mott_2002,wessel_alet_04,rigol_batrouni_09}. Experiments
with a few-site resolution \citep{gemelke_situ_2009}, as well as
single-site resolution \citep{bakr_peng_10,sherson_weitenberg_10},
have been able to resolve the site-occupation profiles and reveal
the characteristic ``wedding cake'' structure in which Mott plateaus
are flanked by superfluid domains. This phenomenology, for sufficiently
deep lattices, can be described within the Bose-Hubbard model \citep{fisher_weichman_89,jaksch_bruder_98}.

While adiabatically slow variations of the lattice potential can be
used as a tuning knob for quantum phase transitions in systems of
trapped atoms, quenching that potential can be utilized as a means
of probing the dynamics. Using this approach, in a recent experiment
on quasi-one-dimensional quantum gases in an optical lattice, it was
demonstrated that quasiparticle pairs transport correlations with
a finite velocity across the system, leading to an effective light
cone for the quantum dynamics \citep{cheneau_light-cone-like_2012}.
Another possibility is to quench the harmonic trap \citep{schneider_hacke_12,ronzheimer_schreiber_13,xia_zundel_14}.
It has been recently shown theoretically that a statistical analysis
of the temporal fluctuations in weak quenches can be used to study
phase transitions \citep{campos_venuti_universal_2014,campos_venuti_universality_2010}.
In this paper, we adapt this fluctuation analysis to examine boundaries
between spatially coexisting phases in trapped systems after a quench
of the trapping potential.

In optical lattice experiments in which $U/J$ is not too large, but
larger than the critical value for the formation of Mott insulating
domains, it is challenging to accurately determine the boundaries
between insulating and superfluid regions. This because the Mott insulator
may exhibit sizable fluctuations of the site occupancies so single
shot measurements of the latter in an insulating plateau may not look
all that different from those in the superfluid region close by. For
the purpose of accurately determining the boundaries between those
domains, several local compressibilities have been proposed in the
literature \citep{batrouni_mott_2002,wessel_alet_04,rigol_local_2003,rigol_muramatsu_04},
including $\kappa_{i}:=\partial\langle\hat{n}_{i}\rangle/\partial\mu_{i}$
\citep{batrouni_mott_2002}, as well as the site-occupation fluctuations
$\Delta n_{i}^{2}:=\langle\hat{n}_{i}^{2}\rangle-\langle\hat{n}_{i}\rangle^{2}$
\citep{rigol_local_2003,rigol_muramatsu_04}, where $\langle\bullet\rangle$
stands for the quantum expectation value and $\mu_{i}$ is the local
chemical potential at site $i$.

Here we propose the use of an out-of-equilibrium quantity, the temporal
variance of the expectation values of site occupancies, as a precise
indicator of boundaries between domains. $\mathcal{N}_{i}(t)=\langle\hat{n}_{i}(t)\rangle$
is the expectation value of $\hat{n}_{i}(t)$, the site-occupation
operator at site $i$ and at time $t$ (in the Heisenberg picture).
The temporal variance of this expectation value is given by 
$\Delta\mathcal{N}_{i}^{2}:=\overline{\mathcal{N}_{i}^{2}}-\overline{\mathcal{N}_{i}}^{2}$,
where $\overline{\bullet}$ denotes the infinite-time average 
$\overline{f}=\lim_{T\to\infty}T^{-1}\int_{0}^{T}f(t)dt$.
Out-of-equilibrium dynamics can be triggered by making a small, sudden
change in the confining potential or the lattice depth. After such
a change (referred to as a quench), the site occupation expectation
values $\mathcal{N}_{i}(t)$ oscillate in time. Our numerical analysis
of the temporal variance of $\mathcal{N}_{i}(t)$, and of the compressibility
$\kappa_{i}$, shows that $\Delta\mathcal{N}_{i}^{2}$ has several
features that make it attractive as an indicator of spatial phase
boundaries. Specifically, when compared to $\kappa_{i}$, (i) the
temporal variance shows a stronger divergence with system size at
the boundary between domains, i.e., $\Delta\mathcal{N}_{i}^{2}\propto L^{\alpha}$
with an exponent $\alpha$ which is larger than that for $\kappa_{i}$
($L$ is the linear system size); and (ii) $\Delta\mathcal{N}_{i}^{2}$
detects finer details in the occupation profile, which are not resolved
by $\kappa_{i}$. The scaling of these quantities with system size
is motivated by analytical results obtained for homogeneous systems.
The scaling analysis also emphasizes the point that beyond a certain
system size, the temporal variance is strictly larger than the local
compressibility.

Furthermore, we discuss an experimentally feasible way to study temporal
fluctuations, based on a small number of temporal sampling points.
We also show that a detailed analysis of the full temporal distribution
$P_{\mathcal{N}_{i}}$ of $\mathcal{N}_{i}$ reveals that deep in
the incompressible region, $P_{\mathcal{N}_{i}}$ is a single peaked,
approximately Gaussian, narrow distribution, whereas in the boundaries
with the superfluid part $P_{\mathcal{N}_{i}}$ is a double-peaked
function indicating bistability and absence of equilibration. We should
stress that, in this work, by small quenches we mean that the system
after the quench needs to be sufficiently close to the initial equilibrium
state, so that time fluctuations of site occupancies are not exponentially
small as one would expect them to be in global quenches in generic
systems \citep{rigol_thermalization_2008}.

The exposition is organized as follows. In Sec.~II, we recapitulate
results for homogeneous systems and present an overview of the temporal
variance $\Delta\mathcal{N}_{i}^{2}$ and of the compressibility $\kappa_{i}$.
In Sec.~III, we apply the proposed technique for identifying phase
boundaries to (integrable) hard-core boson systems. We also investigate
scaling properties of the variance, as those systems allow us to obtain
exact results for very large lattice sizes. We extend this analysis
to a (nonintegrable) $J$-$V$-$V'$ system in Sec.~IV, and comment
on the experimental viability of this approach in Sec.~V. Finally,
we present our conclusions in Sec.~VI.

\section{Quenches and Observables}

Temporal fluctuations following a quantum quench have been studied
extensively in the context of homogeneous systems \citep{noneq_dynamics_review}.
Since some of these results form the motivation for our analysis of
inhomogeneous systems, we briefly review relevant prior work. We consider
systems initialized in the ground state of a Hamiltonian $\hat{H}_{0}=\sum_{n}E_{n}|n\rangle\langle n|$.
The quantum quench is then performed by suddenly changing the Hamiltonian
to $\hat{H}=\hat{H}_{0}+\delta\lambda\,\hat{B}$. For definiteness,
we assume that the perturbation $\hat{B}$ is local and extensive,
i.e.,~$\hat{B}=\sum_{i}\hat{B}_{i}$ with $\left\Vert \hat{B}_{i}\right\Vert =O(L^{0})$
in the system size $L$, where $i$ denotes sites in a lattice. At
time $t$ after the quench, the system's state is given by 
$|\psi(t)\rangle=\exp(-it\hat{H})|\psi(0)\rangle$
(setting $\hbar=1$). For quenches with $\delta\lambda=O(L^{0})$
and a generic observable $\hat{A}$, the expectation value 
$\mathcal{A}(t)=\langle\hat{A}(t)\rangle=\langle\psi(t)|\hat{A}|\psi(t)\rangle$
oscillates around an average value with fluctuations 
$\Delta\mathcal{A}^{2}=\overline{\langle\hat{A}(t)\rangle^{2}}-\overline{\langle\hat{A}(t)\rangle}^{2}$
that are exponentially small in the system volume %
\footnote{By volume $V$ we mean the total volume normalized to the unit cell,
i.e.,~the number of elementary cells%
} (see, e.g., Ref.~\citep{campos_venuti_gaussian_2013}). In other
words, $\Delta\mathcal{A}^{2}=\mathcal{O}(e^{-\alpha V})$, where
$\alpha$ is a positive constant. However, if the quench amplitude
$\delta\lambda$ is comparatively small {[}i.e.,~$\delta\lambda\sim O(L^{-q})$
for some exponent $q>0$ to be specified{]}, the original state is
not completely destroyed during the post quench time evolution. As
a result, such quench experiments can be used to obtain information
on the pre-quench state of the system.

As shown in Ref.~\citep{campos_venuti_universal_2014}, the temporal
variance for such small quenches is of order $\delta\lambda^{2}$,
and is given by 
\begin{equation}
\Delta_{B}\mathcal{A}^{2}=2\delta\lambda^{2}\sum_{n>0}\left|Z_{n}\right|^{2}+O\left(\delta\lambda^{3}\right),\label{eq:var_AB}
\end{equation}
with $Z_{n}:=A_{0,n}B_{n,0}/\left(E_{n}-E_{0}\right)$ and the notation
$A_{n,m}=\langle n|\hat{A}|m\rangle$. The subscript $B$ in $\Delta_{B}\mathcal{A}^{2}$
indicates that the variance is computed for time evolution following
a quench $\delta\lambda\,\hat{B}$. A simple condition for neglecting
the cubic term in Eq.~(\ref{eq:var_AB}) can be written as $\delta\lambda^{2}\chi_{F}\ll1$,
where $\chi_{F}$ is the fidelity susceptibility \citep{zanardi_ground_2006}.
Using the scaling law in Ref.~\citep{campos_venuti_quantum_2007},
one obtains the condition $\delta\lambda\ll\min\{L^{-d/2},L^{-1/\nu}\}$,
where $\nu$ is the correlation length critical exponent.

Equation~\eqref{eq:var_AB} shows an intriguing similarity to the
zero temperature equilibrium isothermal susceptibility $\chi_{AB}$
defined by $\langle\psi\left(\delta\lambda\right)|\hat{A}|\psi\left(\delta\lambda\right)\rangle=\langle\psi\left(0\right)|\hat{A}|\psi\left(0\right)\rangle-\delta\lambda\,\chi_{AB}+O\left(\delta\lambda^{2}\right)$,
where $|\psi\left(\delta\lambda\right)\rangle$ is the ground state
of $\hat{H}=\hat{H}_{0}+\delta\lambda\,\hat{B}$. Indeed, we have
\begin{equation}
\chi_{AB}=2\sum_{n>0}\mathrm{Re}Z_{n}.
\end{equation}

Moreover, using Eq.~(\ref{eq:var_AB}), we see that, to second order
in $\delta\lambda$, we have $\Delta_{B}\mathcal{A}^{2}=\Delta_{A}\mathcal{B}^{2}$,
where we define $\mathcal{B}(t):=\langle\hat{B}(t)\rangle=\langle\psi(t)|\hat{B}|\psi(t)\rangle$.
The same duality holds for the susceptibility, i.e.,~$\chi_{AB}=\chi_{BA}$
for Hermitian operators $\hat{A},\,\hat{B}$.

For systems with a non-zero spectral gap $\Delta$, one can further
relate susceptibilities to quantum fluctuations. One can show that
$\chi_{AA}\le(2/\Delta)\Delta A^{2}$, where $\Delta A^{2}=\langle\hat{A}^{2}\rangle-\langle\hat{A}\rangle^{2}$
is the (zero-temperature) quantum fluctuation of $\hat{A}$.

Both, quantum fluctuations and generalized susceptibilities, are commonly
used indicators of critical behavior in homogeneous systems. Here,
we advocate for temporal fluctuations as a superior indicator. For
homogeneous systems (and extensive observables) one can show that
all these quantities are extensive in gapped, non-critical, systems
\citep{campos_venuti_unitary_2010,campos_venuti_universal_2014}.
Instead, in the critical region (defined by $\xi\gg L$) one can use
scaling hypothesis to predict that the behavior at criticality is
\citep{campos_venuti_universal_2014}: 
\begin{align}
\Delta A^{2} & \sim L^{2d-2\Delta_{A}}\label{eq:scaling1}\\
\chi_{AB} & \sim L^{q},\quad q=2d+\zeta-\Delta_{A}-\Delta_{B}\label{eq:scaling2}\\
\Delta_{B}\mathcal{A}^{2} & \sim L^{2q}\,.\label{eq:scaling3}
\end{align}

\begin{figure}
\vspace{0.2cm}

\begin{centering}
\includegraphics[clip,width=8cm]{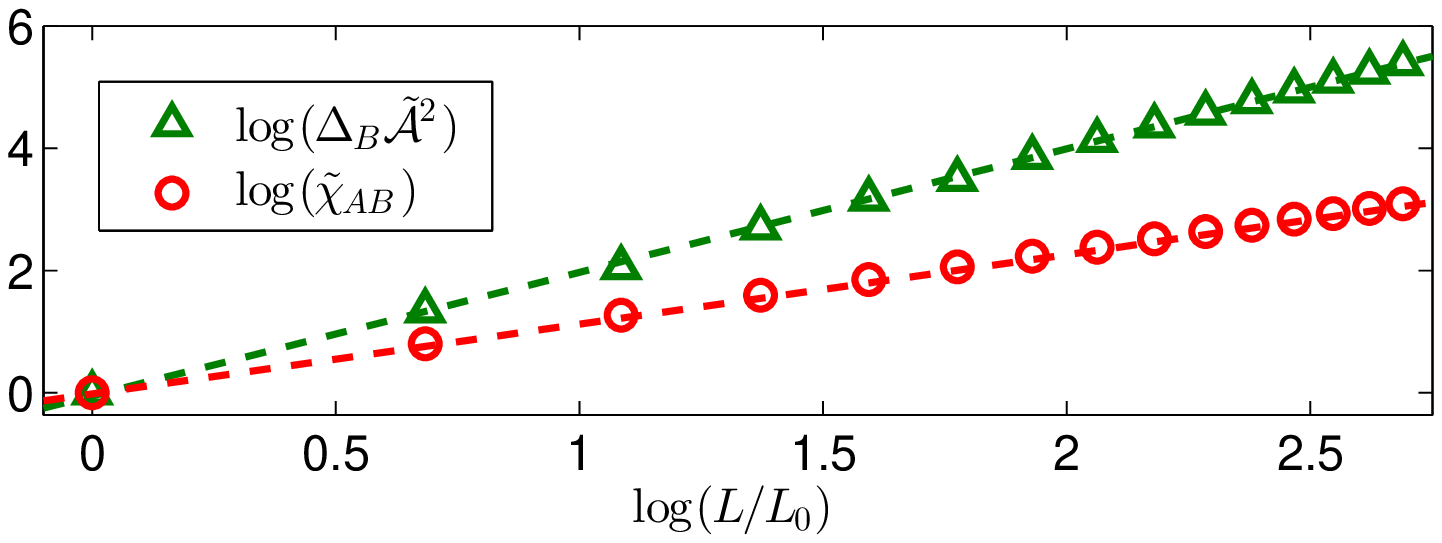} 
\par\end{centering}

\vspace{-0.4cm}
 \protect\protect\protect\protect\protect\caption{(Color online) Verification of Eqs.~(\ref{eq:scaling2}) and (\ref{eq:scaling3}),
which predict $\chi_{AB}\propto L$ and $\Delta_{B}\mathcal{A}^{2}\propto L^{2}$
when $q=1$. A best fit to algebraic scaling gives $\Delta_{B}\mathcal{A}^{2}\propto L^{2.0}$
and $\chi_{AB}\propto L^{1.1}$. We consider a tight binding model
at half filling with both the observable and perturbation set to $\hat{A}=\hat{B}=\sum_{i}(-1)^{i}\hat{n}_{i}$.
$\Delta_{B}\mathcal{A}^{2}$ and $\chi_{AB}$ are made dimensionless
by dividing by their values at $L_{0}=102$, i.e., we plot $\widetilde{\chi}_{\mathrm{AB}}=\chi_{\mathrm{AB}}(L)/\chi_{\mathrm{AB}}(L_{0})$
and $\Delta_{B}\widetilde{\mathcal{A}}^{2}=\Delta_{B}\widetilde{\mathcal{A}}^{2}(L)/\Delta_{B}\widetilde{\mathcal{A}}^{2}(L_{0})$.
\label{fig:notrap_scaling}}
\end{figure}

Note that if the perturbation $\hat{B}$ is relevant, one has $d+\zeta-\Delta_{B}=1/\nu>0$
and the exponent $q$ can be written as $q=d+1/\nu-\Delta_{A}$. The
above equations make it clear that the strongest divergence is exhibited
by the temporal fluctuations. For simplicity, setting $\hat{A}=\hat{B}$,
the exponents satisfy $2d-2\Delta_{A}<q<2q$.

We verify Eqs.~(\ref{eq:scaling2}) and (\ref{eq:scaling3}) for
a tight binding model of spinless fermions $\hat{H}_{0}=\sum_{i=1}^{L}\left[-J(\hat{f}_{i}^{\dagger}\hat{f}_{i+1}+\hc)\right]$
at half filling (and periodic boundary conditions) and observable/perturbation
given by $\hat{A}=\hat{B}=\sum_{j}(-1)^{i}\hat{n}_{i}$. We find that
the scaling of $\chi_{AB}$ and $\Delta_{B}\mathcal{A}^{2}$ is in
accordance with the predictions of Eqs.~(\ref{eq:scaling2}) and
(\ref{eq:scaling3}) with $q=1$ (see Fig.~\ref{fig:notrap_scaling}).

Before we discuss inhomogeneous systems, which are the ones relevant
in the context of ultracold atom experiments, we would like to clarify
how the local compressibility $\kappa_{i}$ relates to the observables
introduced so far. Straightforwardly, we realize that $\kappa_{i}=\chi_{AA}$
with $\hat{A}=\hat{n}_{i}$, whereas $\mathcal{N}_{i}(t):=\langle\hat{n}_{i}(t)\rangle$
and its (temporal) variance $\Delta_{B}\mathcal{N}_{i}^{2}$ are the
dynamical counterparts. We focus on a small perturbation of the strength
of a trapping potential of the form $\hat{B}=L^{-2}\sum_{i}(i-i_{0})^{2}\hat{n}_{i}$.
Here $i_{0}$ is the location of the trap center, and the potential
is normalized by $L^{2}$ to ensure extensivity of the Hamiltonian.
(In principle, other quenches are clearly possible in which the system
is perturbed by varying different parameters, e.g.,~the lattice depth.)
Advocating a local density approximation, one can assume that a very
large trapped system can be divided into extensive regions, where
each region can be considered approximately homogeneous. In this case,
the scaling prediction for the temporal variance of the site occupations,
is 
\[
\frac{\Delta_{B}\mathcal{N}_{i}^{2}}{(\delta\lambda)^{2}}\sim\begin{cases}
O(1) & i\,\mathrm{in\, the\, gapped\, region}\\
L^{\gamma} & i\,\mathrm{in\, the\, critical\, region}
\end{cases}\,,
\]
with a new scaling exponent $\gamma$. According to Eqs.~(\ref{eq:scaling1})--(\ref{eq:scaling3}),
we expect this exponent $\gamma$ to be larger than the corresponding
ones for the compressibility and site-occupation fluctuations.

\section{Hard-core boson systems}

As a first example for the proposed analysis, let us explore how to
detect spatial boundaries between coexisting phases in an integrable
model, where we can perform numerical simulations for very large systems.
This also allows us to perform a finite-size scaling analysis to compare
the divergence of the temporal fluctuations with that of the compressibility,
demonstrating that the temporal variance exhibits a stronger divergence
at the boundary between domains.

We examine a quantum system of hard-core bosons in one dimension described
by the Hamiltonian 
\begin{equation}
\hat{H}_{0}=-J\sum_{i=1}^{L-1}(\hat{b}_{i}^{\dagger}\hat{b}_{i+1}+\hc)+\lambda\sum_{i=1}^{L}g_{i}\hat{n}_{i},\label{eq:H_tightbind}
\end{equation}
which can be thought of as the limit $U/J\rightarrow\infty$ of the
Bose-Hubbard model \citep{cazalilla_citro_review_11}. In Eq.~\eqref{eq:H_tightbind},
$\hat{b}_{i}^{\dagger}$ ($\hat{b}_{i}$) is the creation (annihilation)
operator of a hard-core boson at site $i$, $\hat{n}_{i}=\hat{b}_{i}^{\dagger}\hat{b}_{i}$,
and $g_{i}$ describes a harmonic confining potential, with $g_{i}=L^{-2}(i-L/2+\epsilon)^{2}$.
The trap is shifted off-center by a small amount $\epsilon$ to remove
degeneracies in the energy levels and gaps of the Hamiltonian {[}see
the discussion of Eq.~\eqref{eq:qfree-var}{]}. We initialize the
system in a ground state $|\Psi(0)\rangle$ of a lattice with $L$
sites and $N$ hard-core bosons. After performing a sudden quench
on the trap potential, $\lambda\rightarrow\lambda+\delta\lambda$
at time $t=0$, the system evolves unitarily as $|\Psi(t)\rangle=\exp(-i\hat{H}t)|\Psi(0)\rangle$.
The post-quench Hamiltonian is given by $\hat{H}=\hat{H}_{0}+\delta\lambda\,\hat{B}$.
The hard-core boson Hamiltonian \eqref{eq:H_tightbind} can be mapped
onto a Hamiltonian quadratic in fermion operators $\hat{f}_{i}^{\dagger}$
and $\hat{f}_{i}$ through the Jordan-Wigner transformation \citep{cazalilla_citro_review_11}.
From that transformation, it follows that the site occupations of
hard-core bosons and spinless fermions are identical. The fermionic
Hamiltonian can be written as $\hat{H}=\sum_{i,j}\hat{f}_{i}^{\dagger}M_{i,j}\hat{f}_{j}$
with $M_{i,j}=-J(\delta_{i,j+1}+\delta_{i,j-1})+(\lambda+\delta\lambda)g_{i}\delta_{i,j}$.
The noninteracting character of the latter system allows one to write
temporal fluctuations of site occupations (and in fact of any quadratic
observable in the fermions) in terms of one-particle quantities alone.
Consider the general observable $\hat{X}=\sum_{i,j}\hat{f}_{i}^{\dagger}\Gamma_{i,j}\hat{f}_{j}$.
One can show that $\langle\Psi(t)|\hat{X}|\Psi(t)\rangle=\mathcal{X}(t)=\tr(\hat{X}e^{-it\hat{H}'}\hat{\rho}_{0}e^{it\hat{H}'})=\tr(\Gamma e^{-itM}Re^{itM})$
where $R$ is the covariance matrix of the initial state $\hat{\rho}_{0},$
i.e.,~$R_{i,j}=\tr(\hat{\rho}_{0}\hat{f}_{j}^{\dagger}\hat{f}_{i}^ {})$
(note that the initial state does not necessarily need to be Gaussian).
Let the one-particle Hamiltonian $M$ have the spectral representation
$M=\sum_{k}\Lambda_{k}|k\rangle\langle k|$ ($|k\rangle$ are one
the particle eigenfunctions). Defining $F_{k,q}=\langle k|\Gamma|q\rangle\langle q|R|k\rangle$
where $\Gamma,\, R$ are one-particle operators, the temporal variance
of $\mathcal{X}$ is then given by 
\begin{equation}
\Delta\mathcal{X}^{2}=\sum_{k,q}F_{k,q}F_{q,k}-\sum_{k}(F_{k,k})^{2}.\label{eq:qfree-var}
\end{equation}
Note that Eq.~(\ref{eq:var_AB}) holds for sufficiently small $\delta\lambda$
and relies on the assumption of a non-degenerate many-body spectrum.
Equation~(\ref{eq:qfree-var}), on the other hand, relies on the
assumption of non-resonant conditions for the one-particle spectrum
\citep{reimann_foundation_2008,campos_venuti_gaussian_2013}, which
has been verified in our numerical calculations (for $\epsilon\neq0$).
To compute the variance of the site occupations, we take $\mathcal{X}=\mathcal{N}_{i}$
with $\Gamma_{x,y}^{(i)}=\delta_{i,x}\delta_{i,y}$.

\begin{figure}
\begin{centering}
\includegraphics[clip,width=8cm]{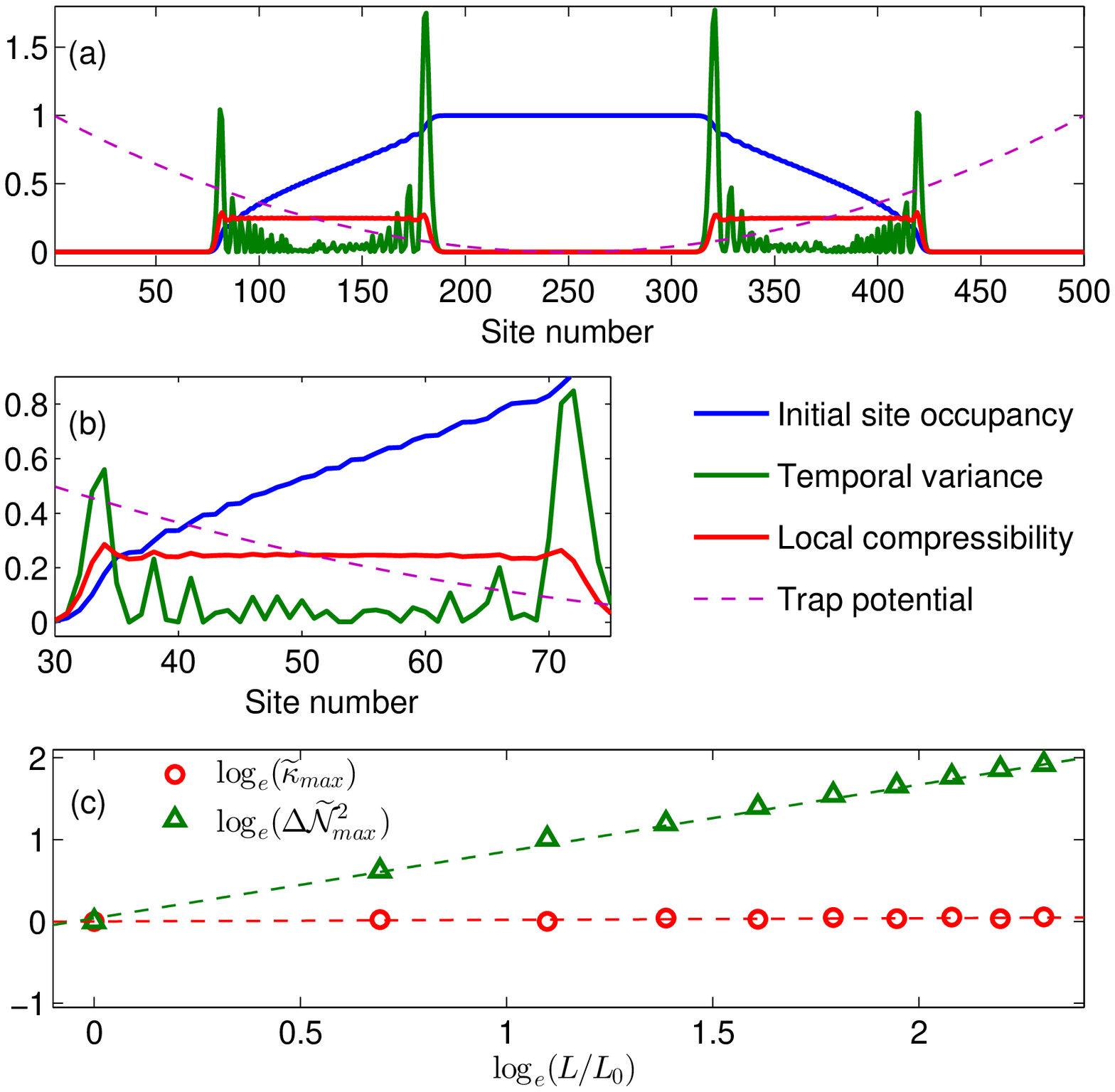} 
\par\end{centering}

\vspace{0.2cm}

\begin{centering}
\includegraphics[clip,width=8cm]{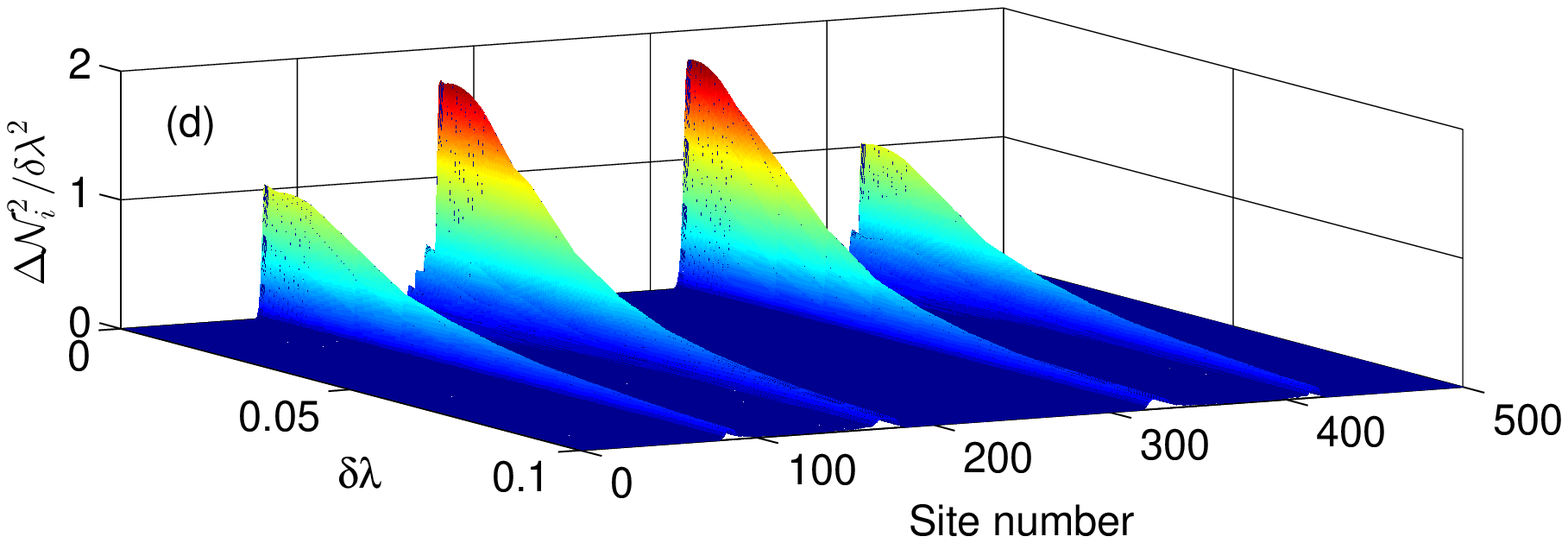} 
\par\end{centering}

\protect\protect\protect\protect\caption{(Color online) (a) ``Wedding cake'' site occupation profile of hard-core
bosons in a one-dimensional harmonic trap described by Eq.~\eqref{eq:H_tightbind}.
The system consists of $L=500$ sites and $N=250$. The Hamiltonian
parameters are $\lambda=10$, $\epsilon=0.2$, $\delta\lambda=L^{-2}$
($J=1$ throughout). The phase boundaries between the Mott plateau
located at the trap center and the adjacent superfluid regions can
be detected by the conventional local compressibility $\kappa_{i}$
(red) and by the temporal variance of the site occupations $\Delta\mathcal{N}_{i}^{2}$
(green) introduced in this work. (b) A closer look at the superfluid
region for the system shown in (a) reveals temporal variance peaks
at the interface between the superfluid and the Mott insulator. (c)
Finite-size scaling of the maximum temporal variance of the site occupations
and of the compressibility vs $L$ for the Hamiltonian in Eq.~(\ref{eq:H_tightbind}).
We find $\Delta\mathcal{N}_{\mathrm{max}}^{2}\propto L^{0.83}$ and
$\kappa_{\mathrm{max}}\propto L^{0.05}$. Both quantities in this
plot are made dimensionless by dividing by their values at $L_{0}=50$,
i.e., $\widetilde{\kappa}_{\mathrm{max}}=\kappa_{\mathrm{max}}(L)/\kappa_{\mathrm{max}}(L_{0})$
and $\Delta\widetilde{\mathcal{N}}_{\mathrm{max}}^{2}=\Delta\mathcal{N}_{\mathrm{max}}^{2}(L)/\Delta\mathcal{N}_{\mathrm{max}}^{2}(L_{0})$.
(d) Dependence of the normalized temporal variance $(\Delta\mathcal{N}_{i}^{2}/\delta\lambda^{2}$)
on the quench amplitude $\delta\lambda$ for the Hamiltonian in Eq.~(\ref{eq:H_tightbind})
with all other parameters as in (a). \label{fig:quasi-free}}
\end{figure}

\begin{figure}
\begin{centering}
\includegraphics[clip,width=8cm]{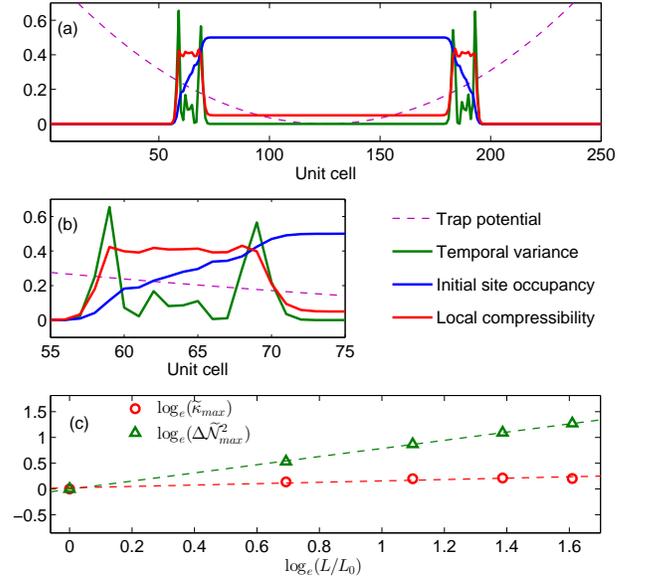} 
\par\end{centering}

\protect\protect\protect\protect\protect\caption{(Color online) (a) Unit cell averaged site occupancy in the presence
of a staggered potential Eq.~\eqref{eq:H_stag}. This is a system
with $L=500$, $N=150$, and parameters $\lambda=10$, $\epsilon=0.2$,
$\delta\lambda=1/L^{2}$, $V_{0}=1.5$. (b) A closer look at the superfluid
region for the system shown in (a) reveals temporal variance peaks
at the interface between the superfluid and the Mott insulator. (c)
Finite-size scaling of the maximum temporal variance of the site occupations
and of the compressibility vs $L$ for the Hamiltonian Eq.~(\ref{eq:H_stag}).
We find $\Delta\mathcal{N}_{\mathrm{max}}^{2}\propto L^{0.80}$ and
$\kappa_{\mathrm{max}}\propto L^{0.14}$. Both quantities in this
plot are made dimensionless by dividing by their values at $L_{0}=50$,
i.e., $\widetilde{\kappa}_{\mathrm{max}}=\kappa_{\mathrm{max}}(L)/\kappa_{\mathrm{max}}(L_{0})$
and $\Delta\widetilde{\mathcal{N}}_{\mathrm{max}}^{2}=\Delta\mathcal{N}_{\mathrm{max}}^{2}(L)/\Delta\mathcal{N}_{\mathrm{max}}^{2}(L_{0})$.\label{fig:quasi_free_stagger}}
\end{figure}

Results of our numerical simulations are shown in Fig.~\ref{fig:quasi-free},
where the site occupations are plotted along with the two measures
of local critical behavior we wish to compare here. Clearly both quantities
are able to distinguish the superfluid regions from the insulating
plateau at the trap center. The local compressibility $\kappa_{i}$
vanishes in the plateau (band insulating) regions where the state
is close to $|1,1,\ldots,1\rangle$ (trap center) and near the trap
boundaries with state $|0,0,\ldots,0\rangle$. Also, $\kappa_{i}$
is roughly constant in the superfluid region. In contrast, $\Delta\mathcal{N}_{i}^{2}$
fluctuates strongly within the superfluid regime, displaying sharp
peaks delineating the insulating regime from its surroundings. A closer
look at the site occupation profiles {[}Fig.~\ref{fig:quasi-free}(b){]}
reveals that, due to the finite size of the system studied (which
will also be the case in experiments), the site occupations at the
boundary between insulating and superfluid domains change in a stepwise
fashion. $\Delta\mathcal{N}_{i}^{2}$ displays clear signatures of
the presence of such steps in the site occupation profiles, while
they are barely reflected in $\kappa_{i}$.

More importantly, a finite-size scaling analysis reveals that the
maxima of $\Delta\mathcal{N}_{i}^{2}$ diverge much more rapidly with
system size than the maxima of $\kappa_{i}$ . We first verified that
the size $\ell$ of the intermediate region between the two band insulator
states scales as $\ell\sim L/c$ with $c\approx4$. A fit to numerical
data {[}see Fig.~\ref{fig:quasi-free}(d){]} reveals power-law dependencies
on system size $L$ (or equivalently, on $\ell$) 
\begin{align}
\Delta\mathcal{N}_{\mathrm{max}}^{2}\propto L^{0.83}\\
\kappa_{\mathrm{max}}\propto L^{0.05}
\end{align}
The scaling seen in Fig.~\ref{fig:quasi-free}(d) makes apparent
that, beyond some system size (that will depend on the Hamiltonian
parameters), the signal given by $\Delta\mathcal{N}_{i}^{2}$ will
exceed that of $\kappa_{i}$. This means that the boundaries between
domains can be determined with higher confidence using the temporal
measure, provided the systems are not too small.

In general, insulating states realized in experiments exhibit nonzero
quantum fluctuations of the site occupancies. This is to be contrasted
to the quantum fluctuations of the site occupancies in the band insulating
phases of Hamiltonian \eqref{eq:H_tightbind}, which are always zero.
In order to address what happens in the presence of nonzero quantum
fluctuations of the site occupancies, while still retaining the advantages
of dealing with models mappable to noninteracting ones, we add a staggered
potential to Eq.~\eqref{eq:H_tightbind} and consider 
\begin{equation}
\hat{H}_{0}=-J\sum_{i=1}^{L-1}(\hat{b}_{i}^{\dagger}\hat{b}_{i+1}+\hc)+\sum_{i=1}^{L}[\lambda g_{i}\hat{n}_{i}+V_{0}(-1)^{i}\hat{n}_{i}].\label{eq:H_stag}
\end{equation}
The properties of systems which such a Hamiltonian have been previously
studied for spinless fermions \citep{rigol_confinement_2004} and
hard-core bosons \citep{rigol_muramatsu_06}. The ground state displays
site-occupation fluctuations within the insulating phase with average
site occupancy of 1/2. Those fluctuations vanish as $V_{0}\rightarrow\infty$,
in which case the insulator becomes a product state of the form $|0,1,0,\ldots,1,0\rangle$.
Accordingly, we plot all quantities in Fig.~\ref{fig:quasi_free_stagger}(a)
averaged over (two site) unit cells. As seen in Fig.~\ref{fig:quasi_free_stagger}(a),
for this model and the parameters chosen, the insulating plateau in
the center of the trap is larger relative to the size of the superfluid
domains than the one in the absence of the staggered potential. Nonetheless,
the superfluid domains are clearly identifiable using $\Delta\mathcal{N}_{i}^{2}$
and $\kappa_{i}$ (notice that $\kappa_{i}$ is nonzero also in the
insulating domain). Studying the finite-size scaling of the maximum
of both quantities, we find the temporal variance and compressibility
scaling to be, 
\begin{align}
\Delta\mathcal{N}_{\mathrm{max}}^{2}\propto L^{0.80}\\
\kappa_{\mathrm{max}}\propto L^{0.14}
\end{align}
{[}see Fig.~\ref{fig:quasi_free_stagger}(c){]}, respectively. Therefore
the same conclusions hold regarding better detectability of spatial
phase boundaries using $\Delta\mathcal{N}_{i}^{2}$ for sufficiently
large system sizes. Interestingly, we have found that the systems
sizes for which $\Delta\mathcal{N}_{i}^{2}$ starts to give a stronger
signal than $\kappa_{i}$ are larger in the presence than in the absence
of the staggered potential. This results from having nonzero charge
fluctuations in the insulator in the center of the trap.

The dependence of the normalized temporal variance $(\Delta\mathcal{N}_{i}^{2}/\delta\lambda^{2}$)
on the quench amplitude $\delta\lambda$ is depicted in Fig.~\ref{fig:quasi-free}(c)
for the Hamiltonian in Eq.~\eqref{eq:H_tightbind}. As expected,
the normalized temporal variance decays for increasing $\delta\lambda$,
following a linear regime $(\Delta\mathcal{N}_{i}^{2}\propto\delta\lambda^{2}$)
for small quenches ($\delta\lambda<1/L^{2}$).

\section{Nonintegrable systems}

In order to show that the proposed approach works beyond integrable
Hamiltonians such as the ones analyzed in the previous section, here
we consider a nonintegrable model. We should stress that the exponential
increase of the Hilbert space with system size severely restricts
the system sizes that can be studied numerically. We focus on a system
consisting of hard-core bosons with nearest and next-nearest interactions
(a $J$-$V$-$V'$ model) in the presence of a harmonic trap, described
by the Hamiltonian 
\begin{align}
\hat{H} & =\sum_{i=1}^{L-1}\left[-J(\hat{b}_{i}^{\dagger}\hat{b}_{i+1}+\hc)+V\left(\hat{n}_{i}-\frac{1}{2}\right)\left(\hat{n}_{i+1}-\frac{1}{2}\right)\right.\nonumber \\
 & \left.+V'\left(\hat{n}_{i}-\frac{1}{2}\right)\left(\hat{n}_{i+2}-\frac{1}{2}\right)\right]+\lambda\sum_{i=1}^{L}i^{2}\hat{n}_{i}.\label{eq:J-V-V1}
\end{align}
Note that, in order to maximize the size of insulating and superfluid
domains, in Eq.~(\ref{eq:J-V-V1}) we only consider one half of what
would be the harmonic trap in an experiment.

\begin{figure}
\begin{centering}
\includegraphics[clip,width=8cm]{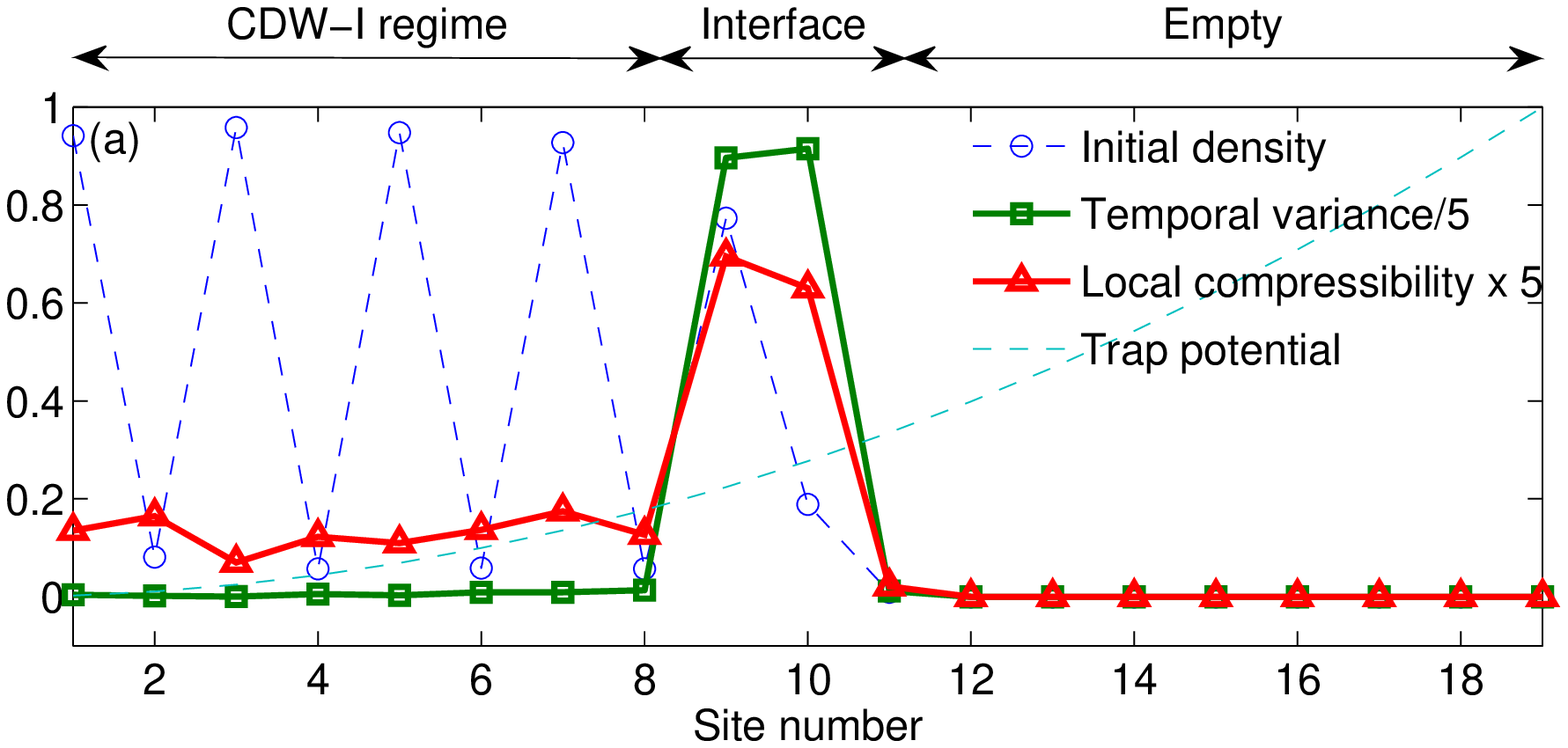} 
\par\end{centering}

\vspace*{0.2cm}

\begin{centering}
\includegraphics[width=8cm]{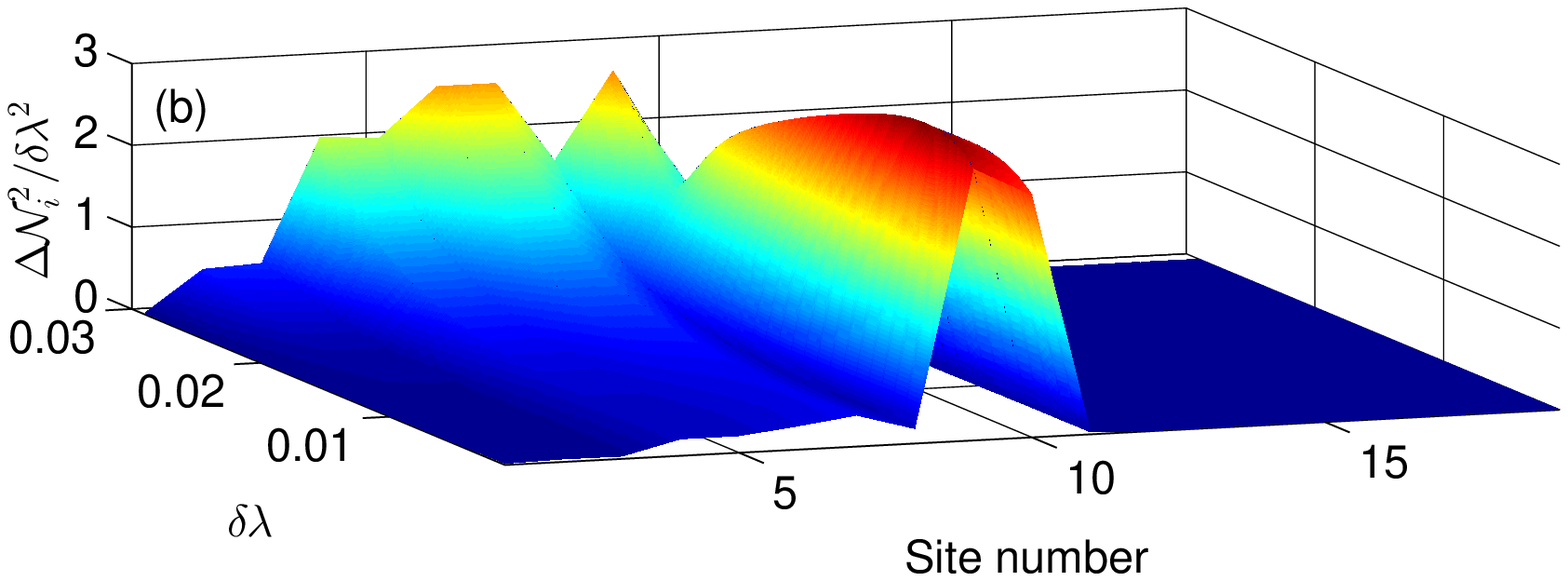} 
\par\end{centering}

\protect\protect\protect\protect\protect\caption{(Color online) (a) Spatial profile of the temporal variance of the
site occupations $\Delta\mathcal{N}_{i}^{2}$ and of the local compressibility
$\kappa_{i}$ for the model in Eq.~(\ref{eq:J-V-V1}). We initialize
the system with 19 sites and 5 particles in the ground state with
parameters $J=1$, $V=8.0$, $V'=0.5$ and $\lambda=0.1225$. The
quench is performed by changing the trap potential from $\lambda$
to $\lambda+\delta\lambda$ with $\delta\lambda=0.0061$. (b) Dependence
of the variance on the quench amplitude $\delta\lambda$. \label{fig:non-integrable}}
\end{figure}

In the absence of a trap, the phase diagram of Hamiltonian \eqref{eq:J-V-V1}
has been studied using the density matrix renormalization group technique
\citep{mishra_phase_2011}. The competition between nearest-neighbor
and next-nearest-neighbor interactions generates four phases: two
charge-density-wave insulator phases, a superfluid (Luttinger-liquid)
phase, and a bond-ordered phase. In the presence of a trap, and for
a suitable choice of the parameters, the same four phases can be observed.
We focus our analysis on a parameter regime where the system exhibits
a charge density wave of type one (CDW-I) in the center of the trap,
which is surrounded by a superfluid phase. The site occupations in
the CDW-I phase are similar to those in the presence of the superlattice
potential analyzed in the previous section, when the average site
occupation per unit cell is 1/2 (see Fig.~\ref{fig:non-integrable}).
In contrast to the superlattice case, the CDW-I phase here is not
due to the presence of a translationally symmetry breaking term but
is stabilized by the presence of interactions. There are two other
phases that have larger unit cells, consisting of 4 sites for CDW-II
and 3 sites for bond-order. The CDW-I phase is the best suited for
our purposes because we are able to observe several unit cells that
exhibit its expected properties.

In Fig.~\ref{fig:non-integrable}(a), we show results for a site-occupation
profile exhibiting a CDW-I plateau surrounded by a small superfluid
domain. In the same figure one can see that, at the edge of the CDW-I
plateau, the local compressibility $\kappa_{i}$ exhibits a much weaker
signal than the temporal fluctuations $\Delta\mathcal{N}_{i}^{2}$.
(Note that we used multiplicative factors to enhance $\kappa_{i}$
and reduce $\Delta\mathcal{N}_{i}^{2}$ so that both measures can
appear on the same scale). Also, notice that $\kappa_{i}$ does not
vanish in the CDW-I plateau, which exhibits nonzero site occupation
fluctuations. Since calculations for larger systems are prohibitively
large, a finite-size scaling analysis of the observables is not possible
here. Nonetheless, from Fig.~\ref{fig:non-integrable}(a), it is
evident that the temporal variance is a better indicator of the interface
between domains than the local compressibility. In fact, compared
to the integrable systems considered in the preceding section, the
advantage of using $\Delta\mathcal{N}_{i}^{2}$ over $\kappa_{i}$
to identify interfaces between domains is enhanced, especially taking
into account the small system sizes considered here.

In Fig.~\ref{fig:non-integrable}(b), we plot $\Delta\mathcal{N}_{i}^{2}/\delta\lambda^{2}$
vs $\delta\lambda$. Similarly to the results in the previous section,
we notice a decrease in the peak height with increasing $\delta\lambda$,
following a linear regime $(\Delta\mathcal{N}_{i}^{2}\propto\delta\lambda^{2}$)
at small $\delta\lambda$. For $\delta\lambda\gtrsim0.2$, a qualitatively
different behavior sets in. This is because the CDW-I domain is destroyed
by the final trap and an $n=1$ Mott insulating domain appears at
the potential minimum of the trap. The latter domain gives rise to
a large temporal variance of the site occupations at its end, which
is located in sites that were formerly in the CDW-I regime.

\begin{figure}
\begin{centering}
\includegraphics[clip,width=8cm]{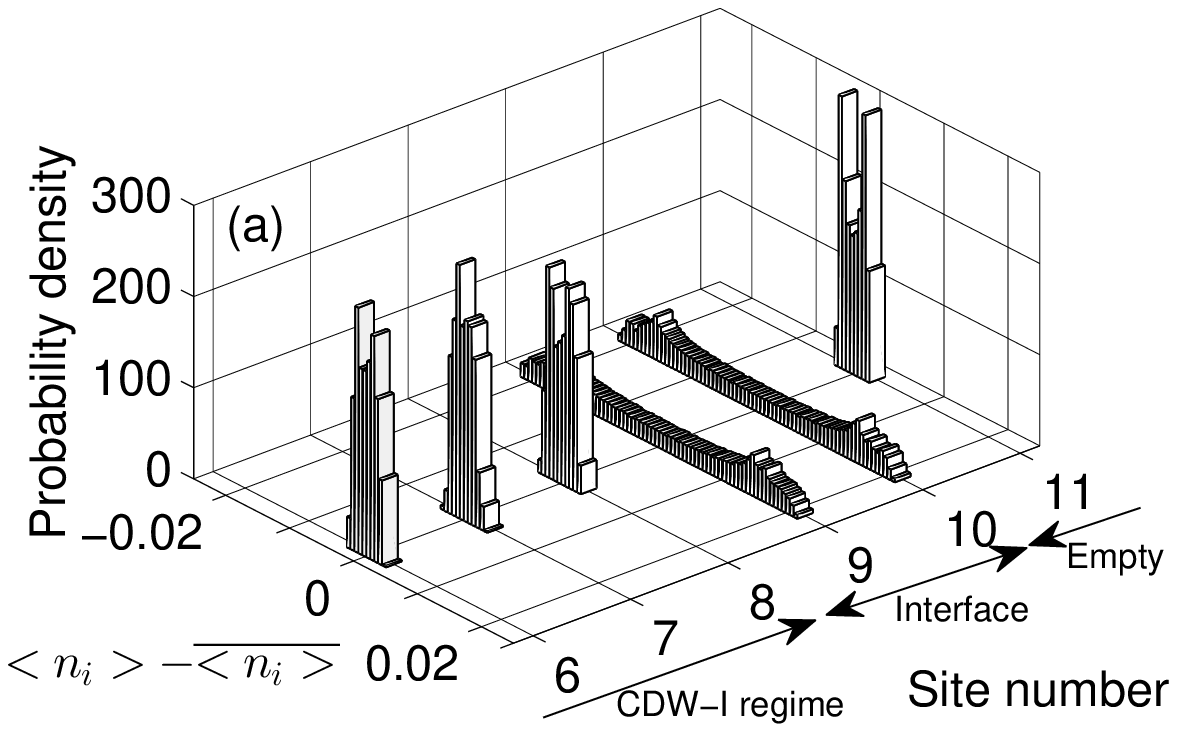} 
\par\end{centering}

\vspace{0.5cm}

\begin{centering}
\includegraphics[clip,width=8cm]{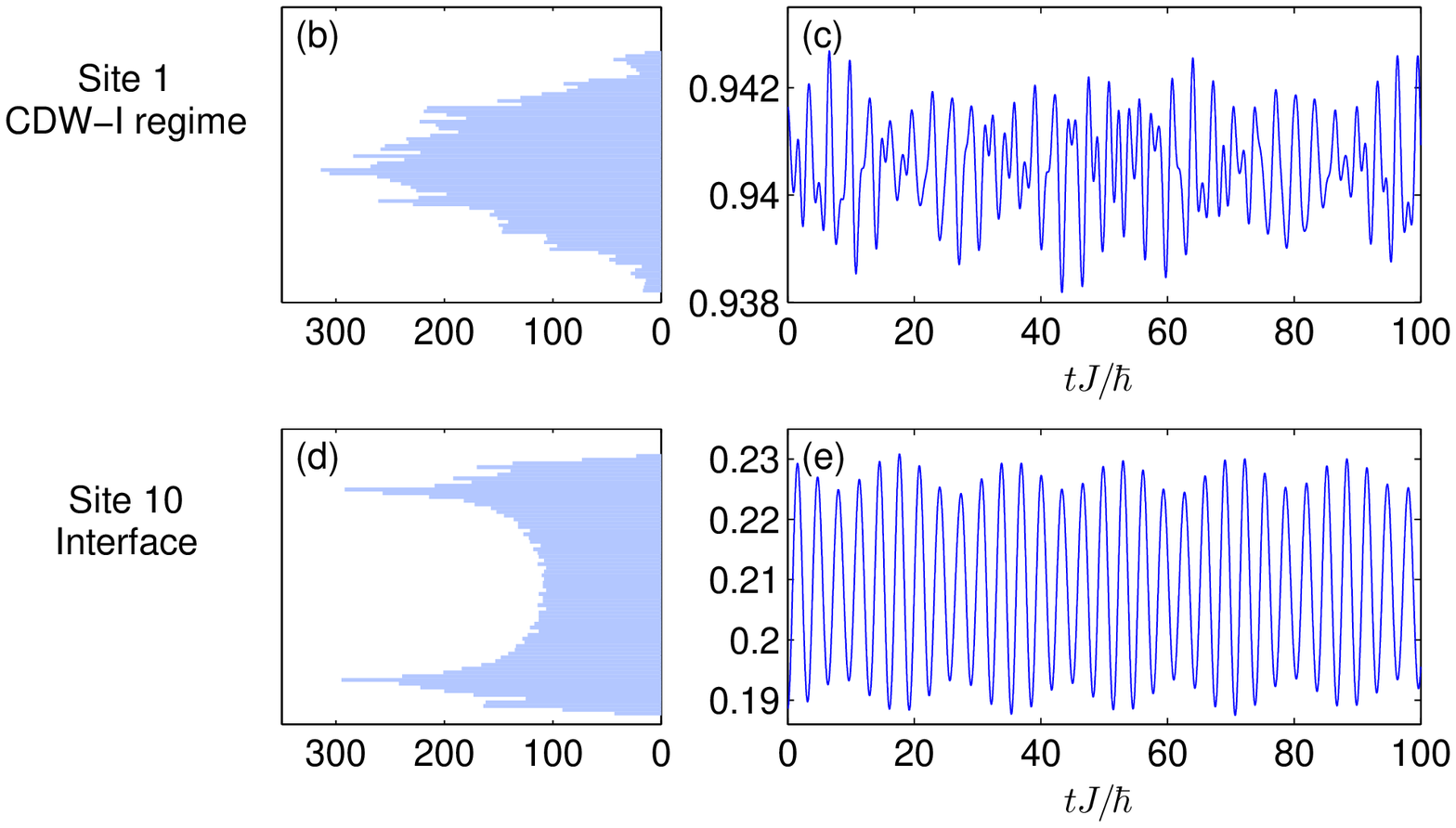} 
\par\end{centering}

\protect\protect\protect\protect\protect\caption{(Color online) (a) Distributions of the site occupations $\mathcal{N}_{i}(t)$
at sites near the interface between the CDW-I and the superfluid phase.
(b) and (d) Distribution function of the site occupation at a site
deep in the CDW-I regime (site $i=1$) and at a site at the edge of
the CDW-I domain (site $i=10$), respectively. (c) and (e) Time dependence
of $\mathcal{N}_{i}(t)$ corresponding to (b) and (d), respectively.
These results are obtained from simulations with the same Hamiltonian
and system parameters as in Fig.~\ref{fig:non-integrable}. Each
$\mathcal{N}_{i}(t)$ is sampled at $N=4\times10^{4}$ random times
uniformly distributed in $\left[0,T\right]$ with $T=40\hbar/J$.
\label{fig:full_distributions}}
\end{figure}

We now go beyond the second moment analysis presented so far and examine
the full probability distribution $P_{i}(x)$ of the random variable
$\mathcal{N}_{i}(t)$ equipped with the time average measure $\overline{\bullet}$.
Based on the results for homogeneous systems \citep{campos_venuti_universality_2010,campos_venuti_universal_2014},
we expect $P_{i}(x)$ to be a single peaked, approximately Gaussian,
narrow distribution for sites $i$ deep in the (gapped) insulating
regime. On the contrary, $P_{i}(x)$ is predicted to be a double peaked
distribution with a relatively large variance for (critical) interface
sites $i$. In a limiting, somewhat simplified case, $P_{i}(x)$ can
be approximated by a two parameter distribution $P_{i}(x)=1/\left(\pi\sqrt{2\Delta\mathcal{N}_{i}^{2}-(x-\overline{\mathcal{N}_{i}})^{2}}\right)$
\citep{campos_venuti_universality_2010}.

In Fig.~\ref{fig:full_distributions}(a), we show the distribution
$P_{i}(x)$ for sites near the interface separating the insulating
and superfluid regions. For sites $i$ deep in the insulating region
{[}Fig.~\ref{fig:full_distributions}(b){]}, the site occupations
fluctuate about one unique central value, resulting in a singly-peaked
distribution function. This signifies measure concentration, indicating
local equilibration in the finite system considered here {[}Fig.~\ref{fig:full_distributions}(c){]}.
In contrast, as one moves closer to the interface {[}Fig.~\ref{fig:full_distributions}(d){]},
the expectation values of observables can be approximated as \citep{campos_venuti_universality_2010}
\begin{equation}
\mathcal{A}(t)\simeq\overline{\mathcal{A}(t)}+\mathcal{A}_{1}\cos[(E_{1}-E_{0})t]+\mathcal{A}_{2}\cos[(E_{2}-E_{0})t]+\ldots\label{eq:A_boundary}
\end{equation}
with the remaining terms being negligible (the constants $\mathcal{A}_{1,2}$
depend on the initial state, evolution Hamiltonian and first excited
states, see \citep{campos_venuti_universality_2010} for details).
The probability distribution then develops peaks at $\overline{\mathcal{A}(t)}\pm\vert\vert\mathcal{A}_{1}\vert-\vert\mathcal{A}_{2}\vert\vert$
and a relatively large variance \citep{campos_venuti_universality_2010}.
This bistability indicates a lack of measure concentration and a breakdown
of local equilibration {[}see Fig.~\ref{fig:full_distributions}(e){]}.

\section{Measuring temporal variances}

So far we assumed that the expectation value $\mathcal{A}(t_{j})=\langle A(t_{j})\rangle$
could be determined exactly for various times $t_{j}$. In this section,
we take a deeper look at the issue of estimating the temporal variance
$\Delta\mathcal{A}^{2}$ using measurement data, keeping in mind ultracold
atom experiments. In these experiments, one typically obtains information
about site occupations by taking a ``snapshot'' of the system
\citep{bakr_peng_10,sherson_weitenberg_10} at a given time $t_{j}$
after the quench. In that case, the observable of interest
is the on-site occupation number. We keep our discussion general so 
that it can be applied to any observable. The expectation value 
$\mathcal{A}(t_{j})=\langle A(t_{j})\rangle$
can be estimated by performing $N_{S}$ measurements of $A$ after
the \emph{same} amount of time $t_{j}$ after the quench. When $A$
 is compactly supported {[}for Fermions $\hat{n}_{i}(t_{j})$ is actually
Bernoulli distributed{]} the error in estimating $\mathcal{A}(t_{j})$
decreases exponentially with $N_{S}$ as a consequence of the Chernoff
bound. One strategy to estimate the temporal variance would be to
take $N_{S}$ sufficiently large such that $\mathcal{A}(t_{j})$ can
be obtained with the desired precision. One then needs to repeat the
above procedure at $N_{T}$ different times $\{t_{1},t_{2},\ldots,t_{N_{T}}\}$
where $t_{i}\in[0,T]$ to estimate the temporal variance%
\footnote{The question about how large should be $T$ to have 
$\Delta\mathcal{A}_{T}^{2}\approx\Delta\mathcal{A}^{2}$
has been addressed in \citep{campos_venuti_equilibration_2013}. Typically
taking $T$ to be one or two revival times gives very accurate results.
The revival time is the time required by excitations to travel the
length of the system, and is hence proportional to the linear size
$L$ \citep{campos_venuti_equilibration_2013}. The proportionality
constant is a time-scale which is of the order of the tunneling time
$\hbar/J$. %
}. As a result, a total of $N_{S}N_{T}$ measurement are required (in
principle $N_{s}$ may depend on $j$, but we do not consider this
generalization here). However this may not be the best strategy to
obtain the temporal variance $\Delta\mathcal{A}^{2}$. In practice
one wants to minimize the total number of measurements.

In order to design better strategies, we look deeper into the measurement
problem in our out-of-equilibrium setting. We recall it here for clarity:
the system is prepared, $N_{S}N_{T}$ times, in the same initial state
$\rho_{0}$ at time $t=0$ and allowed to evolve unitarily thereafter
with the same Hamiltonian parameters. Let us denote with $A_{p}(t_{j})$
the result of the $p$-th measurement of $A$ performed at time $t_{j}$,
$p=1,\ldots,N_{S}$, $j=1,\ldots,N_{T}$ (i.e., one of the eigenvalues
of $A$). The random variables $A_{p}(t_{j})$ at different times
are independent but not identically distributed (as opposed to measurements
performed in equilibrium, in which case they are identically distributed).

In the language of statistics, what we would like to build is a \emph{consistent
estimator} of the temporal variance $\Delta\mathcal{A}^{2}$. A consistent
estimator is a method to obtain a given quantity with the property
that, as the number of data point increases, the estimator converges
to the actual parameter we are trying to estimate (see
e.g.~Ref.~\citep{amemiya_1985}). In our case the data points are
the random variables $A_{p}(t_{j})$. The quantum expectation value
$\mathcal{A}(t_{j})=\langle A(t_{j})\rangle$ is estimated using $N_{S}$
measurements by
\[
e_{j}=\frac{1}{N_{S}}\sum_{p=1}^{N_{S}}A_{p}(t_{j}),
\]
which converges to $\mathcal{A}(t_{j})$ in the large $N_{S}$ limit.
We now define the following estimator $v$ for the the temporal variance
of $A$: 
\begin{equation}
v=\frac{1}{N_{T}}\sum_{j=1}^{N_{T}}(e_{j}-\mu)^{2}\quad\textrm{with }\quad\mu=
\frac{1}{N_{T}}\sum_{j=1}^{N_{T}}e_{j}.\label{eq:s}
\end{equation}
Using $\mathsf{E}[\bullet]$ to denote expectation value over all
the $N_{S}N_{T}$ independent measurements, we find,
\begin{align*}
\mathsf{E}[v] & =\frac{1}{N_{T}N_{S}}\left\{ (N_{S}-1)
\sum_{j}\langle A(t_{j})\rangle^{2}+\sum_{j}\langle A^{2}(t_{j})\rangle\right\} \\
 & -\frac{1}{N_{T}^{2}}\sum_{j\neq k}\langle A(t_{j})\rangle\langle A(t_{k})\rangle\\
 & -\frac{1}{N_{T}^{2}N_{S}}\left\{ (N_{S}-1)\sum_{j}\langle A(t_{j})\rangle^{2}+
 \sum_{j}\langle A^{2}(t_{j})\rangle\right\} .
\end{align*}

\begin{figure}
\begin{centering}
\includegraphics[width=8cm]{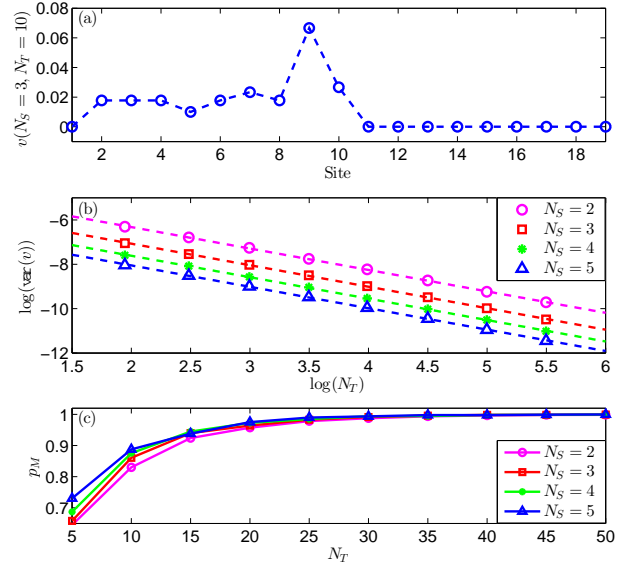} 
\par\end{centering}
\protect\protect\protect\protect\protect\caption{(Color online) Estimating the 
temporal variance. (a) An example of
the temporal variance estimator $v$ at each site for $N_{S}=3$ and
$N_{T}=10$. All parameters are the same as for Fig.~\ref{fig:non-integrable}.
We compute $v$ independently for each of the $L=19$ sites of the
system. (b) Scaling of the variance of the estimator $v$ at the site
$i=10$ with $N_{T}$, for different values of $N_{S}$. The fit shows
$\mathsf{var}(s_{i})\sim N_{T}^{-0.97}$ for the $N_{S}$ values considered,
in accordance with our prediction that $v$ is a consistent estimator.
(c) Results of a numerical experiment to compute $p_{M}$, the probability
that $\{v_{i}\}$ and the exact variance have maxima at the same sites
$(i=9,10)$. $10^{4}$ samples of $\{v_{i}\}$ were used to compute
these probabilities. \label{fig:estimator_s}}
\end{figure}

We still have to specify how to choose the $N_{T}$ times. If we pick
the times randomly with uniform distribution in $[0,T]$ and denote
with$\mathsf{T}[\bullet]$ the corresponding time average operation
(i.e.,~$\mathsf{T}$ averages uniformly over all $N_{T}$ independent
times $t_{j}$), we obtain
\begin{align*}
&\mathsf{T}[\mathsf{E}[v]]  =\left[\frac{N_{S}-1}{N_{S}}\overline{\left(\langle A(t)
\rangle^{2}\right)^{T}}-\frac{N_{T}-1}{N_{T}}\overline{\langle A\rangle^{T}}^{2}\right]\\
 & +\frac{1}{N_{S}N_{T}}\left[(N_{T}-1)\overline{\langle A(t)^{2}\rangle^{T}}-(N_{S}-1)
 \overline{\left(\langle A(t)\rangle^{2}\right)^{T}}\right],
\end{align*}
where we indicated $\overline{f(t)^{T}}=T^{-1}\int_{0}^{T}f(t)dt$.
We see that in the limit $N_{S},N_{T}\rightarrow\infty$, the expectation
value of this estimator tends to the exact variance $\overline{(\mathcal{A}^{2})^{T}}-
\left(\overline{\mathcal{A}{}^{T}}\right)^{2}=:\Delta\mathcal{A}_{T}^{2}$.
This means that $v$ is an asymptotically unbiased\textcolor{black}{{}
estimator}, i.e.,~when the number of measurement increases the estimator
converges to the exact temporal variance. Furthermore, we have numerically
checked that $v$\textcolor{black}{{} is }also a \emph{consistent}
estimator, meaning that the error on $v$, encoded in 
$\mathsf{var}[v]=\mathsf{T}[\mathsf{E}[v^{2}]]-\mathsf{T}[\mathsf{E}[v]]^{2}$,
tends to zero as the number of measurements increases. In Fig.~\ref{fig:estimator_s}(b), 
we show that $\mathsf{var}[v]\sim N_{T}^{-1}$. 

We now show the feasibility of this approach for distinguishing different
domains in trapped systems. We perform numerical experiments on the
Hamiltonian in Eq.~\eqref{eq:J-V-V1}. According to our general recipe,
we perform a small quench of the trapping potential and measure the
occupation number at each site during the following time evolution.
In this case, the observable is the site occupation $A=\hat{b}_{i}^{\dagger}\hat{b}_{i}$
and we use $v_{i}$ to denote the corresponding temporal variance
estimated according to Eq.~(\ref{eq:s}) for $i=1,\ldots,L$.
As mentioned earlier, $A_{p}(t_{j})$ is the result of the $p$-th
measurement of $A$ at time $t_{j}$. In our numerical experiments,
this is obtained by randomly generating one of the eigenvalues of
$A$ ($0$ or $1$ for $A=\hat{b}_{i}^{\dagger}\hat{b}_{i}$) with
probabilities given by the Born rule.

In figure \ref{fig:estimator_s}(a), we show a typical realization
of $v_{i}$ obtained taking $N_{T}=10$ and $N_{S}=3$,
for a total of $30$ measurements. One can compare Fig.~\ref{fig:estimator_s}
with Fig.~\ref{fig:non-integrable}(a), where the exact variance
is plotted for the same parameters. Clearly, the maxima at sites $9,10$
in Fig.~\ref{fig:estimator_s}(a) predicts a transition
region in agreement with that in Fig.~\ref{fig:non-integrable}(a). 
Still, we are primarily interested in the
efficacy of $v_{i}$ in locating the boundary
between coexisting phases. In other words, we are interested in knowing
whether the position of the maxima of $v_{i}$ coincides with that
of the exact variance [sites 9 and 10 as seen in in Figs.~\ref{fig:non-integrable}
and \ref{fig:estimator_s}(a)]. To this end, we compute the 
probability $p_{M}(N_{T})=\mathrm{Prob}(\arg\max_{i}\{v_{i}\}\in\{9,10\})$
as a function of the number of measurements $N_{T}$. $p_{M}(N_{T})$
is plotted in Fig.~\ref{fig:estimator_s}(c) as a function of $N_{T}$
for different $N_{s}$. We observe that the estimator defined in Eq.~\eqref{eq:s}
allows us to locate the boundary with a $90\%$ accuracy, using
a total of around 40 measurements. These findings suggest that temporal
fluctuations can be used to efficiently locate critical boundaries in
experiments. 

\section{Conclusions}

We have studied various trapped systems whose ground states exhibit
coexistence of insulating and superfluid domains, as relevant to ultracold
atom experiments in optical lattices. An analysis of the time evolution
of the site occupations $\mathcal{N}_{i}$, following a small quench
of the trapping potential, allowed us to show that the temporal variance
of $\mathcal{N}_{i}$ can be used as an accurate tool to locate boundaries
between domains. We found that the temporal variance of $\mathcal{N}_{i}$
at those boundaries exhibits a power law scaling with system size
with an exponent that is greater than the one of a previously proposed
local compressibility.

Furthermore, we performed a binning analysis to explicitly study the
\emph{temporal} probability distribution of site occupancies. Such
a temporal distribution gives the probability of observing a given
value of the site occupation in a large observation time-window $[0,T]$.
We observed that the distributions are sharply peaked and approximately
Gaussian for sites that are deep in the insulating phase, while sites
at the interface display a bimodal distribution, i.e., are characterized
by a lack of measure concentration. We further analyzed the feasibility
of our approach from an experimental point of view. We found that
since we are interested in general features of the temporal variance
profile (presence of peaks at phase interfaces), rather than the exact
statistics of $\mathcal{N}_{i}$, sample variances obtained using
small number of measurements (around 40 for a system with $19$ sites
and $5$ particles) are sufficient for locating phase boundaries 
with high probability. 

\begin{acknowledgments}
The numerical computations were carried out on the University of Southern
California High Performance Supercomputer Cluster. This research was
supported by the ARO MURI grant W911NF-11-1-0268, DOE Grant Number
ER46240, and the US office of Naval Research. S.~Haas would like
to thank the Humboldt Foundation for support. 
\end{acknowledgments}

\bibliographystyle{unsrt}
\bibliography{quench_traps_recovered}

\end{document}